\documentclass[]{aa} 

\usepackage{graphicx}
\usepackage{txfonts}
\usepackage{xcolor}
\usepackage{subcaption}         
\usepackage{placeins}           

\usepackage[breaklinks, colorlinks, citecolor=blue, linkcolor=blue]{hyperref}
\usepackage{listings}

\definecolor{codegreen}{rgb}{0,0.6,0}
\definecolor{codegray}{rgb}{0.5,0.5,0.5}
\definecolor{codepurple}{rgb}{0.58,0,0.82}
\definecolor{backcolour}{rgb}{0.95,0.95,0.92}

\lstdefinestyle{mystyle}{
    backgroundcolor=\color{backcolour},   
    commentstyle=\color{codegreen},
    keywordstyle=\color{magenta},
    numberstyle=\tiny\color{codegray},
    stringstyle=\color{codepurple},
    basicstyle=\ttfamily\scriptsize,
    breakatwhitespace=false,         
    breaklines=true,                 
    captionpos=b,                    
    keepspaces=true,                 
    showspaces=false,                
    showstringspaces=false,
    showtabs=false,                  
    tabsize=2
}

\DeclareMathOperator*{\argmin}{arg\,min}

\newcommand{\vect}[1]{\boldsymbol{\mathbf{#1}}}

\newcommand{\urlcode}{\url{http://github.com/aasensio/inv_emulator}}

\begin{document} 

    \title{Very fast 2D inversions of Stokes profiles using neural emulators}
    
    \author{A. Asensio Ramos\inst{1,2}, J. de la Cruz Rodr\'{\i}guez\inst{3}, C. Kuckein\inst{1,2}, S. J. Gonz\'alez Manrique\inst{1,2} 
    \and C. D\'{\i}az Baso\inst{4,5} }
    
    \institute{
    Instituto de Astrof\'isica de Canarias (IAC), Avda V\'ia L\'actea S/N,
    38200 La Laguna, Tenerife, Spain\\
    \email{andres.asensio@iac.es}
    \and
    Departamento de Astrof\'isica, Universidad de La Laguna, 38205 La Laguna, Tenerife, Spain    
    \and
    Institute for Solar Physics, Dept. of Astronomy, Stockholm University, AlbaNova University Centre, SE-10691 Stockholm, Sweden
    \and
    Institute of Theoretical Astrophysics,
    University of Oslo, %
    P.O. Box 1029 Blindern, N-0315 Oslo, Norway
    \and
    Rosseland Centre for Solar Physics,
    University of Oslo, %
    P.O. Box 1029 Blindern, N-0315 Oslo, Norway
    }

    \date{Received ; accepted }
    
 
    \abstract
    {}
    {We aim to carry out fast, spatio-temporally regularized inversions of
    full fields of view, optionally including the PSF of the telescope, for both photospheric and
    chromospheric spectral lines.}
    {We replace the forward model by fast, precise, and differentiable neural emulators trained on
    synthetic Stokes profiles. Neural surrogate models are used to emulate the
    Stokes profiles of the Fe\,{\sc i} doublet at 6302~\AA, the Ca\,{\sc ii} line at 8542 \AA\ as well as the Si\,{\sc i} line 
    and He\,{\sc i} multiplet at 10830~\AA. These emulators are then used in a fast
    inversion code that includes spatial regularization terms and can compensate for the
    aberrations of the telescope.}
    {The emulators are unbiased, with errors that are, in most cases, comparable to or below the noise
    of current observations in the polarization profiles. Inversions of large fields-of-view can be carried out in minutes
    in a off-the-shelf GPU, opening up the possibility of real-time analysis of large-scale solar observations. We provide
    a few illustrative examples. Emulators open up the option of inverting lines with very complex and time
    consuming forward models.}
    {}
    
    \titlerunning{Very fast 2D inversions of Stokes profiles using neural emulators}
    \authorrunning{Asensio Ramos et al.}
    \maketitle
    \nolinenumbers

%
\section{Introduction}
\label{sec:introduction}

The availability of spectropolarimetric data has increased dramatically in the last two decades.
Modern facilities such as the Swedish 1-m Solar Telescope
\citep[SST;][]{2003SPIE.4853..341S,2024A&A...685A..32S}, GREGOR
\citep{GREGOR,kleint2020}, Hinode SOT/SP \citep{lites_hinode01,kosugi_hinode07},
and the Daniel K. Inouye Solar Telescope \citep[DKIST;][]{2020SoPh..295..172R}
produce data at a rate that traditional inversion strategies struggle to follow. 
Inverting a single pixel may take seconds in local thermodynamic equilibrium (LTE) and 
minutes in non-LTE, so processing a field-of-view (FoV)
with tens of thousands of pixels can take hours or days even on supercomputers.
This bottleneck has motivated the use of machine learning (ML) to build inversion
tools that deliver results in a fraction of the time
\citep[see][]{2023LRSP...20....4A}. The pioneering work of
\cite{2001A&A...378..316C} trained multilayer perceptrons (MLPs) on synthetic
Stokes profiles to perform Milne-Eddington (ME) inversions almost instantaneously.
\cite{2003NN.....16..355S} applied this strategy to measure magnetic fields in
sunspots, \cite{2005ApJ...621..545S} proposed ways to improve the neural
inversion, and \cite{2008A&A...481L..37C} showed that networks can also recover
depth stratifications from LTE lines.

More recently, the use of ML has expanded to more complex models.
\texttt{SICON} \citep{2019A&A...626A.102A} used
convolutional neural networks (CNNs) to map full Hinode SOT/SP observations to stratified atmospheric
parameters in a matter of milliseconds. Since then, this family of direct inverters has
grown quickly. \texttt{SPIn4D} \citep{2024ApJ...976..204Y} applies a similar
CNN approach to DKIST data, and \texttt{SuperSynthIA}
\citep{2024ApJ...970..168W} produces full-disk vector magnetograms from SDO/HMI
\citep{2012SoPh..275..207S}. One-dimensional CNNs have also been used to accelerate
inversions \citep{2020A&A...644A.129M,2022ApJ...925..176C,2022ApJ...940..147L} and to provide better
initializations for classical codes \citep{2021A&A...651A..31G}, while MLPs
remain useful, for example for finding observational evidence of the hot-wall
effect in magnetic pores \citep{2021A&A...652A..78S}. Transformer architectures
\citep{NIPS2017_3f5ee243} have also been explored \citep{2025A&A...703A..55A,2026ApJ...996...63C}.

A limitation shared by most codes above is that each pixel is inverted
independently, as if a FoV were a set of unrelated 1D problems. This ignores that
neighboring pixels are physically connected and that the telescope point spread function (PSF) mixes
their signals. Two strategies have been proposed to account for this.
Spatially coupled inversions \citep{2012A&A...548A...5V} convolve the synthetic
spectra with the PSF before comparing them with the data, and
\cite{2019A&A...631A.153D} generalized this into a global inversion of all pixels,
even for data of different resolution, through a sparse-matrix formulation and the inclusion of spatial regularization for noise-reducing and problem stabilization purposes.
Generally, the spatial (and temporal) correlation present in the data can be exploited through many different forms of regularization:
for example, \cite{2015A&A...577A.140A} imposed sparsity on the maps of physical parameters
using proximal optimization algorithms, which reduced the effective number of
unknowns, increased the robustness of the solution with respect to pixel-by-pixel
inversions, and allowed PSF compensation. 
More recently, \cite{2025A&A...693A.170D} and \cite{2025ApJ...985L...7J} extended 
this concept to the spatiotemporal domain using neural fields, neural networks that 
represent physical quantities as continuous functions of space and time. 
This approach introduces an implicit regularization that also reduces the number of 
free parameters, increases the effective signal-to-noise ratio, and improves the reconstruction of the physical parameters.
%
However, the computational cost of these iterative, synthesis-based 2D approaches is considerable.
The number of unknowns is the
number of pixels times the number of depth nodes of each parameter; a
$100\times100$ pixel map with 10 depth points for temperature, line-of-sight (LOS) velocity, and
the three Cartesian components of the magnetic field has
$5\times10\times100\times100$ unknowns, i.e., half a million. Every iteration
additionally requires synthesizing all the profiles, which has so far restricted
spatially coupled and regularized inversions mainly to ME or LTE models.
By contrast, convolutional neural networks (CNNs) can also exploit spatial information at a fraction of the computational cost \citep{2019A&A...626A.102A,2024ApJ...976..204Y}, although they typically do not reach the same accuracy as methods that directly fit the Stokes profiles.

Neural networks can also be used in inversion problems with a less dominant role. An interesting route, that we
exploit here, is to replace the forward model itself by a neural emulator, a surrogate model which is fast,
differentiable, and can be trained once for a given spectral line
\citep{2024ApJ...977..101X}. Emulators therefore accelerate the synthesis
of synthetic profiles, potentially enabling complex line formation models, at
the expense of accuracy.
In this paper, we use a neural emulator of the
forward model to carry out 2D
inversions of full FoVs with the spatial regularization approach of
\cite{2015A&A...577A.140A}. Unlike direct neural inverters, the observations are
fitted through a physical forward model, so the inversion is not tied to the predefined
depth structure of the training set. Unlike classical 2D codes, the cost of
synthesis no longer limits the size of the problem, opening up the possibility 
of quickly inverting large FoVs with very complex models.

\section{Emulating Stokes profiles}
\label{sec:emulators}
Any useful emulator needs to fulfill several
requirements. First, it has to be fast, because a 2D inversion of a full FoV requires
millions of evaluations of the forward model. Second, its
output needs to be a smooth function of the inputs so that it is differentiable
with respect to its inputs. This way, the gradient of the merit function with respect to
the physical parameters can be obtained with automatic differentiation. Finally, it has to
be accurate inside the region of the parameter space covered by the training set,
and this region needs to be large enough to describe the observations. Neural networks
fulfill all these requirements if the training set and the representation of the inputs
and outputs are chosen carefully.

Although generally applicable, we will consider in this work emulators for three
spectral regions: i) the pair of photospheric Fe\,{\sc i}
lines at 6301.5 and 6302.5~\AA, as observed routinely by Hinode SOT/SP without
telluric contamination, ii) the
spectral region around 10830~\AA, which contains the photospheric
Si\,{\sc i} line at 10827.1~\AA\ and the chromospheric He\,{\sc i} multiplet at
10830~\AA, and iii) the spectral region around the infrared triplet Ca\,{\sc ii} 8542~\AA\ 
line.

\subsection{Training set for the photospheric and chromospheric lines}
\label{sec:training_set}
Emulators can only be trusted in the region of the parameter space spanned by the
training set, so it needs to be as diverse as possible while still
producing atmospheres that are compatible with those found in the solar atmosphere. To
this end, we follow the strategy of \cite{2025A&A...703A..55A} with some modifications. 
All our model atmospheres are defined by the stratification
of the temperature ($T$), the LOS velocity ($v$), the microturbulent velocity ($v_\mathrm{mic}$),
and the three Cartesian components of the magnetic field vector ($B_x$, $B_y$, $B_z$) in
the reference frame of the LOS. All of them 
are sampled in a common grid of 80 equispaced points in the logarithm
of the optical depth at 500~nm ($\log \tau_{500}$), from $-7.5$ to $1.5$.

The models are built by randomly perturbing the set of semi-empirical models
described in \cite{2025A&A...703A..55A}, covering a wide range of physical conditions 
in the solar atmosphere. All of them are reinterpolated to the common optical depth grid using cubic splines, which are
extrapolated linearly for those models that do not reach the limits of the grid.
Instead of perturbing every single base model, we perturb a convex combination of them.
The weights of the combination are drawn from a Dirichlet distribution with a very
small concentration parameter (0.01). Consequently, most of the samples are dominated
by a single base model, but intermediate atmospheres are also generated with some probability.

Following \cite{2025A&A...703A..55A}, the perturbations are drawn from a Gaussian 
process \citep[GP;][]{Rasmussen2006Gaussian}. If $\vect{x}$ is the vector
containing the values of a given physical quantity at all depths, a new sample is drawn
from the multivariate normal distribution $\mathcal{N}(\bar{\vect{x}},\mathbf{\Sigma})$,
where $\bar{\vect{x}}$ is the specific base model and the covariance
matrix uses a squared exponential kernel:
\begin{equation}
\Sigma_{ij} = \sigma_i \sigma_j \exp \left[ -\frac{(\log \tau_i - \log \tau_j)^2}{\ell^2} \right] + 10^{-8} \sigma_i^2 \delta_{ij},
\end{equation}
where the diagonal term is added as as stabilization. The correlation length $\ell$ is drawn for each model from a log-uniform
distribution in the range $[2,25]$, so that some models are very smooth and others
display structure at relatively small scales (although the perturbation is always continuous). The standard deviation of the perturbation
of the temperature varies linearly from 500~K at $\log \tau_{500}=-7.5$ to 100~K at
$\log \tau_{500}=1.5$. For the rest of parameters, it is constant with height and equal to
5~km~s$^{-1}$ for $v$, 2~km~s$^{-1}$ for $v_\mathrm{mic}$, and 1000~G for the three components
of the magnetic field. Samples that produce temperatures outside $[2500, 10^5]$~K, velocities
outside $[-30,30]$~km~s$^{-1}$, microturbulent velocities outside $[0,5]$~km~s$^{-1}$, or
magnetic field components outside $[-3500,3500]$~G are rejected and drawn again.
Since the semi-empirical models are field-free and static, this procedure produces models
that would rarely contain weak fields and small velocities. To guarantee that this
regime, which dominates large parts of the solar surface, is well represented, the
magnetic field vector and the LOS velocity of each model are multiplied by
independent random factors in the range $[0,1]$ and drawn from a $\mathrm{Beta}(2,3)$, whose
mean is 0.4. In the same spirit, the microturbulent velocity is multiplied
by a factor drawn from $\mathrm{Beta}(2,1)$. The final training set contains $10^6$ models,
while the validation and test sets contain $10^4$ models each.

An emulator conditioned on the full stratification of the LOS velocity would need
to learn how to displace the spectral lines over the entire range of possible Doppler shifts.
This is something that neural networks can do, but it is a waste of capacity, since the
effect of a global shift is trivially described analytically. For this reason, we subtract
from the velocity stratification of each model its value at $\log \tau_{500}=-1$
before the synthesis, so that the emulators only learn the effect of velocity gradients.
The global Doppler shift is reintroduced afterwards, during the synthesis, with the
aid of a simple interpolation.

The Stokes profiles of the Fe\,{\sc i} and Si\,{\sc i} lines are synthesized with \texttt{Hazel2} \citep{2008ApJ...683..542A}, which
uses \texttt{SIR} \citep{1992ApJ...398..375R} to solve the radiative transfer equation for the
photospheric lines in LTE. The electron pressure is obtained assuming hydrostatic
equilibrium, and no macroturbulence is added. The Fe\,{\sc i} doublet is synthesized in 150
equispaced wavelength points between 6300.8921 and 6303.2671~\AA, covering the spectral
range observed by Hinode SOT/SP. The Si\,{\sc i} line is synthesized in 350 equispaced wavelength
points between 10824 and 10833~\AA, which covers the spectral range typically observed by the GRIS \citep{2012AN....333..872C} instrument at GREGOR.
For the Fe\,{\sc i} doublet, we 
train a separate emulator for each line, with the spectral
window split at 6302.0~\AA, in the continuum between the two lines. Splitting the
window reduces the complexity of the function that each emulator has to learn.
Finally, each model is synthesized at a heliocentric angle $\mu=\cos\theta$ randomly chosen in the range $[0.2, 1]$.

Concerning the Ca\,{\sc ii} line, we utilize the RH module included in 
STiC \citep{2001ApJ...557..389U,2019A&A...623A..74D} since this line is formed in non-LTE conditions. 
We used a five bound-level plus ionization continuum model atom that includes the H\&K and 
infrared triplet lines. The RH code solves the statistical equilibrium equations using the 
preconditioning method of \citet{1992A&A...262..209R}. Partial redistribution effects of 
scattered photons were included in the H\&K lines, using the hybrid scheme proposed by 
\citet{2012A&A...543A.109L}. The infrared triplet lines are calculated under the simplifying 
assumption of complete redistribution that is usually considered adequate for these lines 
\citep{1989A&A...213..360U}, also ignoring cross-redistribution effects from Raman scattering.

\subsection{Representation of the stratifications and the Stokes profiles}
\label{sec:representation}
Neural networks are trained much more efficiently if the inputs and outputs
are represented in a compact and well-conditioned form. The stratifications are
smooth functions of optical depth and are oversampled in the 80-point grid.
In order to reduce the dimensionality of the input, we project 
each physical parameter onto a basis of cubic B-splines with 10 coefficients 
on uniformly spaced knots in $\log \tau_{500}$. If $\vect{x}$ is the stratification of
a physical parameter and $\mathbf{B}$ is the $80 \times 10$ design matrix of the B-spline
basis, the coefficients are obtained by least squares, $\vect{c} = \mathbf{B}^\dagger \vect{x}$, where
$\mathbf{B}^\dagger=(\mathbf{B}^T \mathbf{B})^{-1}\mathbf{B}^T$ is the pseudo-inverse of $\mathbf{B}$. The projection matrices are
precomputed and stored inside the emulator, so the emulator still accepts a stratification in the
original grid and remains differentiable with respect to it. The coefficients of each physical parameter
are normalized by dividing them by a fixed constant (5000~K for the temperature, 2~km~s$^{-1}$ for both
velocities, and 500~G for the magnetic field components) so that all inputs
have similar orders of magnitude. The heliocentric angle $\mu$ is passed as an additional input.

The efficient representation of Stokes profiles requires more care. The amplitude of Stokes $Q$, $U$, and $V$ is set by the
magnetic field and spans several orders of magnitude in the training set, from the noise-like
signals of the weak-field regime to the saturated profiles of sunspots. They also
change sign and are exactly zero in the absence of a magnetic field, a property that we
would like to preserve. We find that normalizing the Stokes profiles $Q$, $U$, and $V$ by Stokes $I$ and
compressing them with a symmetric logarithmic transformation works well:
\begin{equation}
\tilde{X}(\lambda) = \mathrm{sign}\left[ \frac{X(\lambda)}{I(\lambda)} \right] \log \left[ 1 + \alpha \left| \frac{X(\lambda)}{I(\lambda)} \right| \right], \quad X \in \{Q,U,V\}.
\label{eq:symlog}
\end{equation}
This transformation is linear for $|X/I| \ll \alpha^{-1}$ and logarithmic for $|X/I| \gg \alpha^{-1}$, so it
compresses the large signals while leaving the weak ones untouched. We use $\alpha=30$ for the Fe\,{\sc i} and Ca\,{\sc ii} lines
and $\alpha=100$ for the Si\,{\sc i} line, which places the transition at polarization degrees of a few percent.
The transformation is exactly invertible, so the emulator
produces Stokes profiles in physical units.

Even after the transformation, the profiles are oversampled in wavelength and
strongly correlated, and their information can be efficiently encoded using
principal component analysis \citep[PCA;][]{2000A&A...355..759R,2005ApJ...620..517S,2005ApJ...622.1265C,2007ApJ...660.1690A}
as a compact basis for the output of the emulators. The PCA basis is computed
independently for each Stokes parameter. We keep the number of
eigenvectors needed to explain a fraction $1-10^{-5}$ of the variance in the training
set. This defines the number of components per Stokes parameter, which are shown in
Table~\ref{tab:emulators}. The polarization profiles need more components than Stokes $I$
because they encode more complex shapes, such as the asymmetries and the
multilobed profiles produced by gradients along the LOS. 
The reconstruction of the profiles from the PCA coefficients is a simple matrix product, which is again differentiable.
Overall, this compresses the input by a factor of ${\sim}8$ and the output by a factor of 2--9, depending on the line (Table~\ref{tab:emulators}).

\begin{table*}
\caption{Summary of the emulators.}
\label{tab:emulators}
\centering
\begin{tabular}{lcccccc}
\hline\hline
Emulator & Spectral range [\AA] & $N_\lambda$ & Inputs & Output components & Architecture & Parameters \\
\hline
Fe\,{\sc i} 6301.5 & 6300.89--6301.99 & 70 & 115 & 22/39/40/36 & Residual MLP, $6 \times 512$ & $6.4 \times 10^6$ \\
Fe\,{\sc i} 6302.5 & 6302.01--6303.27 & 80 & 115 & 20/34/34/32 & Residual MLP, $6 \times 512$ & $6.4 \times 10^6$ \\
Si\,{\sc i} 10827 & 10824--10833 & 350 & 115 & 18/49/49/44 & Residual MLP, $6 \times 512$ & $6.4 \times 10^6$ \\
Ca\,{\sc ii} 8542 & 8539.84--8544.31 & 150 & 115 & 38/55/55/51 & Residual MLP, $6 \times 512$ & $6.4 \times 10^6$ \\
He\,{\sc i} 10830 & 10823--10834 & 256 & 7 & $11 \times 30$ & MLP, $4 \times 512$ & $9.6 \times 10^5$ \\
\hline
\end{tabular}
\tablefoot{The output components refer to the number of principal components for
Stokes $I$, $Q$, $U$, and $V$ for the depth-stratified lines, and for each one of the 11 coefficients
$\epsilon_{I,Q,U,V}$, $\eta_{I,Q,U,V}$, and $\rho_{Q,U,V}$ for He\,{\sc i} (see Sect.~\ref{sec:helium}).
The architecture indicates the number of hidden layers (or residual blocks) and their width.}
\end{table*}

\subsection{Neural architecture and training}
\label{sec:architecture}
Once the inputs and outputs are represented in a compact form, the emulator is reduced to a
regression between two vectors of relatively small dimension. Since the spectral lines are 
sensitive to the gradients of the physical parameters with height, we found that the emulators 
work better if we explicitly provide them.
They can be computed from the B-spline coefficients easily, since the derivative of a B-spline 
of degree $p$ is another B-spline of degree $p-1$. Therefore, we append the 9 coefficients of the derivative 
of each physical parameter to the input B-spline coefficients, after standardizing them with the statistics of 
the training set. This gives a total of 115 inputs: 60 B-spline coefficients (six physical quantities with 10 
B-spline coefficients each) and 54 derivative coefficients, and the heliocentric angle $\mu$.
We used a deep residual MLP \citep{he2016deep} as a regressor. The input is first linearly
embedded in a space of dimension 512, and then passes through six residual blocks
that compute $\vect{h} \leftarrow \vect{h} + \mathbf{W}_2\, \mathrm{SiLU}[\mathbf{W}_1\, \mathrm{LN}(\vect{h})]$, where
LN is a layer normalization \citep{ba2016layer}, SiLU is the Sigmoid Linear Unit activation function \citep{elfwing2018sigmoid}, 
and the hidden layer of each block has dimension 1024.
A final layer normalization and a linear layer produce the PCA coefficients.
Each emulator has ${\sim}6.4 \times 10^6$ parameters.

All emulators are trained by minimizing the mean squared error between the predicted
and the true (standardized) PCA coefficients, with the four Stokes parameters equally weighted\footnote{One can also consider defining the loss function directly for the Stokes parameters, 
instead of the PCA coefficients. Additionally, one can add a loss that enforces the emulation of
the wavelength derivatives, which can potentially improve the precision. We defer this analysis for the future.}.
We used the AdamW optimizer \citep{loshchilov2019decoupled}, a
batch size of 1024, a weight decay of $10^{-5}$, and a learning rate that
grows linearly towards a value of 0.1 during the first five epochs and then follows a cosine decay to zero
\citep{loshchilov2017sgdr} at the end of 300 epochs. In all cases, the model with the
smallest validation loss is the one finally used.

\subsection{The He\,{\sc i} 10830~\AA\ multiplet}
\label{sec:helium}

The simple radiative transfer model used for the He\,{\sc i} multiplet at 10830~\AA\ in \texttt{Hazel2}
suggests a natural way to construct the emulator for this multiplet. The computationally expensive part is the computation of the 11 wavelength-dependent emission ($\epsilon_{I,Q,U,V}$), 
absorption ($\eta_{I,Q,U,V}$), and anomalous dispersion ($\rho_{Q,U,V}$) profiles, which requires solving the statistical equilibrium equations. Once these coefficients are known, the formal solution of
the radiative transfer equation is cheap because the slabs are assumed to be homogeneous. We therefore 
emulate only the coefficients, and compute the emergent Stokes profiles for each slab using 
the formal solution
\begin{equation}
\vect{I} = e^{-\mathbf{K} \Delta s} \vect{I}_0 + \mathbf{K}^{-1} \left( \mathbf{1} - e^{-\mathbf{K} \Delta s} \right) \boldsymbol{\epsilon},
\label{eq:slab}
\end{equation}
where $\vect{I}_0$ is the incident Stokes vector, $\mathbf{K}$ is the propagation matrix built 
from the $\eta$ and $\rho$ coefficients, $\mathbf{1}$ is the identity matrix, and $\Delta s = \tau / \eta_I^\mathrm{max}$ is 
the geometrical thickness of the slab.
This choice has several advantages.
Firstly, the coefficients depend on fewer parameters than the emergent profiles.
The optical depth $\tau$ and the LOS velocity $v$ can be taken into account during the formal solution, the
latter being simply applied with cubic spline interpolation of the emulated coefficients to the Doppler-shifted wavelengths.
The emulator then only needs
to depend on the magnetic field vector, the Doppler width $\Delta v_\mathrm{D}$, the damping parameter $a$, the height $h$, which determines the
anisotropy of the illumination, and the inclination $\theta$ of the LOS. Secondly, emulating 
the coefficients allows us to couple the chromospheric slab with any photospheric model in a consistent way. For the same reason, several slabs can be
placed in series, exactly as \texttt{Hazel2} does when dealing with multiple chromospheric components along the LOS
\citep[e.g.,][]{2020A&A...640A..71K}.

The training set is generated by sampling each Cartesian component of the magnetic field
(in the local vertical reference frame) from a zero-mean normal distribution with a standard deviation of 500~G,
and $\Delta v_\mathrm{D}$, $a$, $h$, and $\theta$ uniformly in the intervals
$[4,18]$~km~s$^{-1}$, $[0,1.5]$, $[0.1,10]$\arcsec, and $[0, 89.9]^\circ$, respectively.
We compute the coefficients for $10^6$ models with \texttt{Hazel2}, in 256 wavelength points between 10823 and 10834~\AA, and an additional
set of $10^4$ models for validation and testing. We normalize the 11 coefficients by $\eta_I^\mathrm{max}$ and the emission coefficients 
by the continuum intensity at disk center \citep{1976asqu.book.....A}. The
number of PCA coefficients used for all outputs are shown in Tab.~\ref{tab:emulators}.
The emulator is the same residual MLP used for emulating the Fe\,{\sc i} lines.

\subsection{Precision of the emulators}
\label{sec:precision}
To quantify the precision of the emulated profiles in physical units
we use 2000 Stokes profiles from a test set at disk center ($\mu=1$), which were never seen
during training. 

\begin{figure*}
\centering
\includegraphics[width=\textwidth]{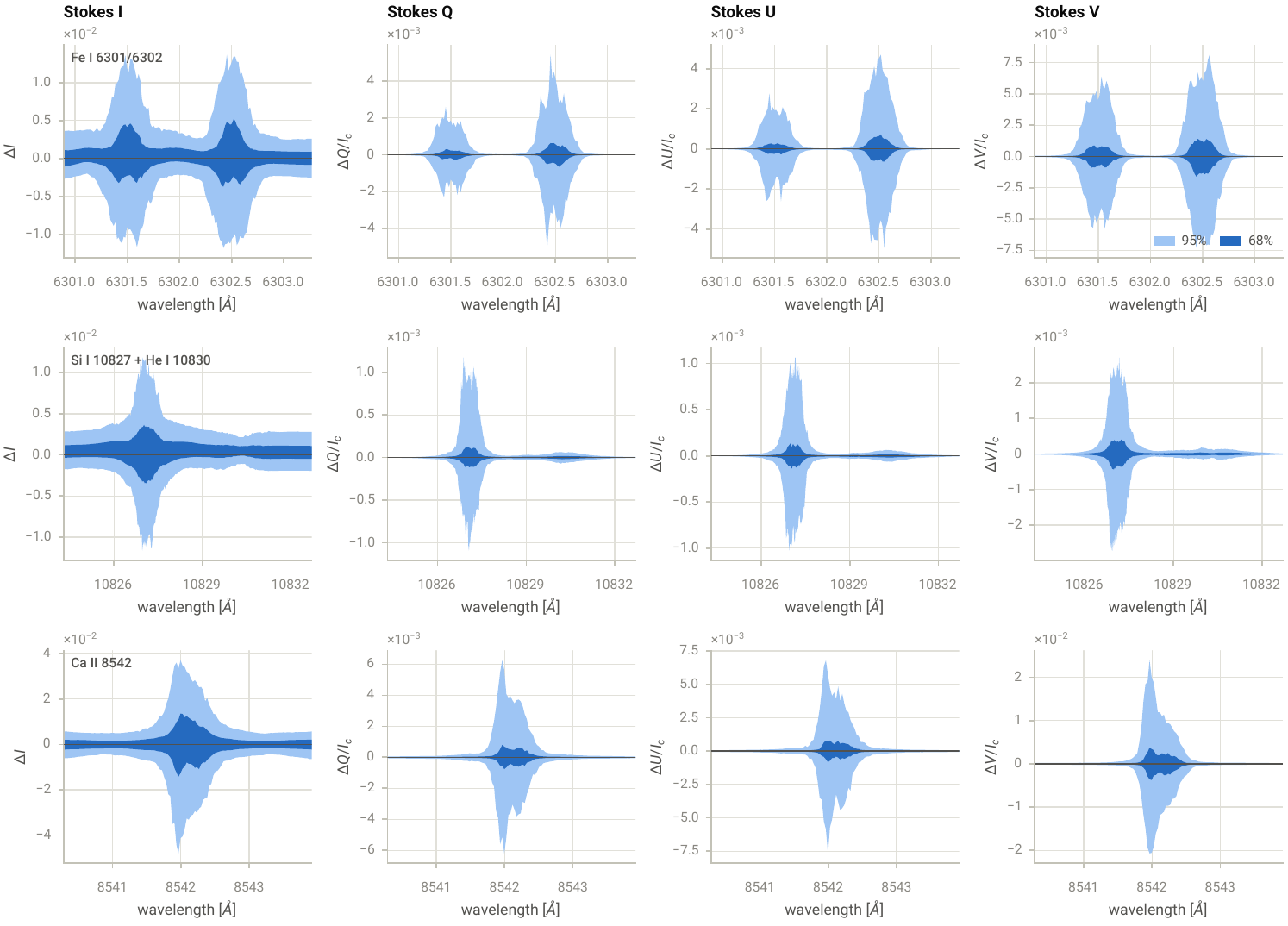}
\caption{Intervals containing 68\% (dark) and 95\% (light) of the differences between the Stokes profiles
synthesized with the emulators and the original profiles for 2000 models of the test set, as a function of wavelength. The upper row
shows the Fe\,{\sc i} doublet at 6301.5 and 6302.5~\AA\, the central row shows the 10830~\AA\ region, including the Si\,{\sc i} line and the He\,{\sc i} multiplet, and the lower row show the results for the Ca\,{\sc ii} line. All differences are in units of the continuum intensity.}
\label{fig:precision_percentiles}
\end{figure*}

Figure~\ref{fig:precision_percentiles} displays the intervals that contain 68\% and 95\% of the
differences between the emulated and the synthetic profiles at each wavelength. All Stokes parameters are
unbiased, with average differences below $5 \times 10^{-4}$ in Stokes $I$ and of the order of $10^{-6}$ in
Stokes $Q$, $U$, and $V$, in units of the continuum intensity. The errors
concentrate at the wavelengths of the spectral lines. For the Fe\,{\sc i}
doublet, 68\% of the differences lie in the interval $\pm 5 \times 10^{-3}$ for Stokes $I$,
in $\pm 7 \times 10^{-4}$ for Stokes $Q$ and $U$, and in $\pm 10^{-3}$ for Stokes $V$, while the 95\% intervals are
$\pm 10^{-2}$, $\pm 5 \times 10^{-3}$, and $\pm 8 \times 10^{-3}$, respectively. The
much wider 95\% intervals point to distributions with extended tails, which are produced by a small
fraction of models (see below). Similar patterns are observed in the 10830~\AA\ region, where the  
errors in this region are dominated by the Si\,{\sc i} line. In the 8542~\AA region, we 
find slightly larger residuals in all Stokes parameters, a direct consequence of the increased complexity
of the line, which is sensitive to a much broader region of the atmosphere.
Despite this difference, we find the emulator precise enough to carry out 
inversions.

\section{Inversions with spatial regularization}
\label{sec:inversions}
Equipped with a fast and differentiable forward model, the inversion of a full FoV can be posed as
a single optimization problem for the physical parameters of all pixels simultaneously.
We use \texttt{PyTorch} \citep{pytorch19} to implement the forward model and use its
automatic differentiation capabilities as well as its parallelization capabilities in CPUs and GPUs to perform 
the inversion. Spatial regularization terms are
added to the merit
function, encoding prior knowledge about the spatial structure of the
maps of physical parameters, as proposed by \cite{2015A&A...577A.140A}.
See App. \ref{sec:appendix_regularization} for details about the currently implemented regularizations.

\subsection{Parameterization of the model atmosphere}
\label{sec:nodes}
The stratification of each physical parameter (except for those of the He\,{\sc i}, which are
constant in the slab)
is described by its values at a few user-selectable nodes in $\log \tau_{500}$. A B-spline
interpolation is used to obtain the stratification in the 80-point depth grid of the emulators.
Consequently, a single node
produces a height-independent parameter, two nodes a linear stratification, and
four or more nodes a cubic spline. 

Since emulators can only be trusted inside the region of the parameter space covered by the
training set, the node values of each parameter, either depth-stratified or constant in the slab, are bounded to the range between
the 0.1 and 99.9 percentiles of the training models after every iteration.
Additionally, we enforce $B_x \geq 0$ for the magnetic field, which removes the
180$^\circ$ ambiguity in the azimuth of the magnetic field vector. Potential 
van Vleck ambiguities in the He\,{\sc i} field are not removed and should be treated in a post-processing step. 

\subsection{Merit function}
\label{sec:merit}
As usual, the inversion minimizes the following merit function:
\begin{equation}
J(\vect{x}) = \sum_{p} \chi^2_p(\vect{x}_p) + \sum_{k} \lambda_k R_k(\vect{X}_k),
\label{eq:merit}
\end{equation}
where $\vect{x}=\{\vect{x}_1, \vect{x}_2, \ldots, \vect{x}_N\}$ contains the concatenation, for all
$N$ pixels in the FoV, of all the vectors $\vect{x}_p$ with the parameters of pixel $p$. Likewise,
$\vect{X}_k$ is the 2D map of
the $k$-th node of a given physical parameter (it is, in essence, a different organization of the
vector $\vect{x}$). The synthetic Stokes profiles can also include the convolution 
with the instrumental profile, both along the spectral and spatial dimensions. 

The data term is the usual $\chi^2$, appropriate for additive noise with diagonal covariance matrix:
\begin{equation}
\chi^2_p = \sum_{s=1}^4 w_s \sum_{i} \left[ \frac{S_s^\mathrm{syn}(\lambda_i;\vect{x}_p) - S_s^\mathrm{obs}(\lambda_i)}{\sigma_s} \right]^2,
\end{equation}
where the sum runs over the four Stokes parameters $S_s$ and over the observed wavelengths $\lambda_i$,
$\sigma_s$ is the noise standard deviation of each Stokes parameter, and $w_s$ are optional weights.
The data term is separable in pixels if the effect of a spatial PSF is not taken into account, so its
gradient can be computed independently for each pixel. As a consequence, without the regularization 
term, optimizing Eq.~(\ref{eq:merit}) pixel by pixel is equivalent to optimizing it for the entire FoV.
The advantage of the latter when using neural emulators is that it allows for a more efficient implementation 
on a GPU, since many pixels can be synthesized in parallel.

The regularization term couples neighboring pixels. 
Each node map can have one or several
regularization functions $R_k$, with weights $\lambda_k$ that can be different for each
physical parameter and each node. We implemented four
regularization functions that the user can choose from (by choosing the weights $\lambda_k$), all of them convex but not necessarily differentiable 
(see App. \ref{sec:appendix_regularization} for a list of regularizers and the technical details of their implementation).

\subsection{Optimization}
\label{sec:optimization}
The optimization of Eq.~(\ref{eq:merit}) is a large-scale, potentially non-differentiable and non-convex problem, with 
millions of unknowns for a typical FoV. Proximal
algorithms \citep{ista09,parikh2013proximal}, that were already used by \cite{2015A&A...577A.140A} for the
same purpose, are the natural tool for this kind of problems. They
alternate a gradient step on the differentiable part of the merit function (the $\chi^2$) with
the application of the proximal operator of the regularization term. See App. \ref{sec:appendix_regularization}
for details on the practical implementation.

\cite{1992ApJ...398..375R} identified that the merit function is strongly sensitive 
to some physical parameters (e.g., temperature) and much less to others (e.g., magnetic field). 
This was solved by using the second order Levenberg-Marquardt algorithm coupled with a 
singular value decomposition of the Hessian matrix. In our case, we use the first order 
Adam optimizer \citep{adam14}, whose diagonal preconditioning adapts the step size to each
unknown. To make the problem better conditioned, the unknowns are the node values divided by
a typical scale of each physical parameter (e.g., 500~K for the temperature, 1~km~s$^{-1}$ for
the velocity, and 200~G for the magnetic field). 
The learning rate of the Adam optimizer starts at $\eta=0.1$ and
follows a cosine law, that decays from its initial value to 5\% of it along the iterations.
The number of iterations is typically set to 1000, although convergence happens much earlier, in
just a few hundred iterations.

From a computational perspective, we note that the memory needed to compute the gradient of 
the $\chi^2$ grows with the number of pixels, since
automatic differentiation has to store the intermediate activations of all the emulators for the backward pass.
If the data term is separable, we split the FoV into chunks of pixels (depending on the size of the 
emulators and the available GPU memory). For each chunk, the
Stokes profiles are synthesized in parallel, the $\chi^2$ is computed and back-propagated, and the computational graph is
released before the next chunk is processed. The gradients with respect to the parameters are accumulated over
the chunks and, once all of them have been processed, the Adam step and the proximal step are applied to the full
maps. Since the gradient of $\chi^2_p$ depends only on the parameters of pixel $p$, the accumulated
gradient is exactly that of the full FoV, so that FoVs of
arbitrary size can be inverted on a single GPU with a peak memory use that is controlled by the chunk size.

The situation changes when the spatial PSF of the telescope is included in the forward model. In this case,
the synthetic Stokes maps of the full FoV, $\vect{S}^\mathrm{syn}$, are convolved with the spatial PSF, $\mathbf{P}$, before
being compared with the observations. Therefore, $S_s^\mathrm{syn}(\lambda_i;\vect{x}_p)$ in the $\chi^2$ is replaced by
$\tilde{S}^\mathrm{syn}_{s,p}(\lambda_i)=[\mathbf{P} * S_s^\mathrm{syn}(\lambda_i)]_p$. The residual of a pixel now 
depends on the parameters of all the
pixels inside the support of the PSF \citep[see, e.g.,][]{2012A&A...548A...5V,2019A&A...631A.153D}, so that the data 
term is no longer separable, and the chunks cannot be processed
independently.
However, the gradient can still be efficiently computed by exploiting the structure of the problem, as shown 
in App. \ref{sec:appendix_psf}. 

\begin{figure*}
\centering
\includegraphics[width=0.8\textwidth]{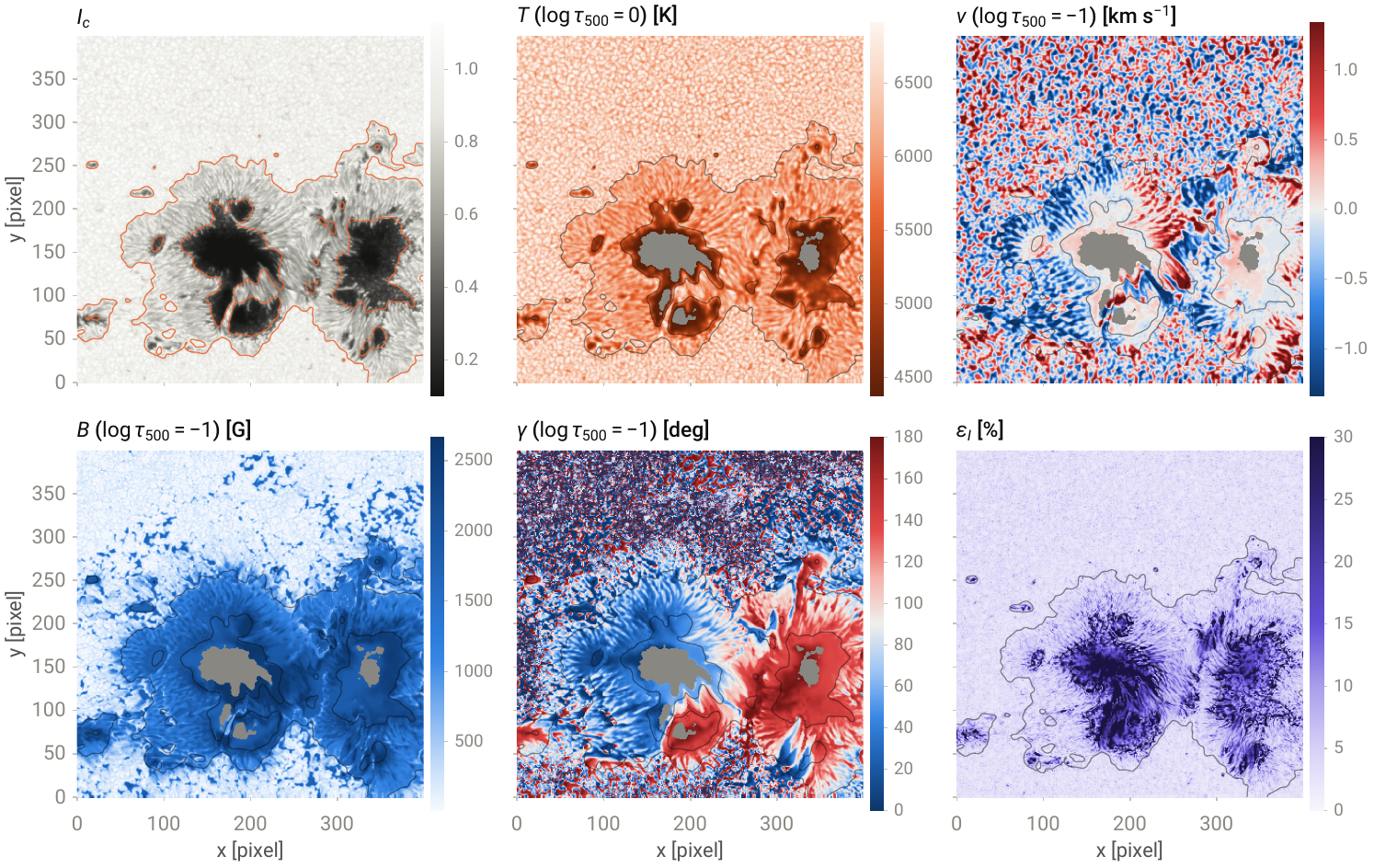} \\
\includegraphics[width=0.8\textwidth]{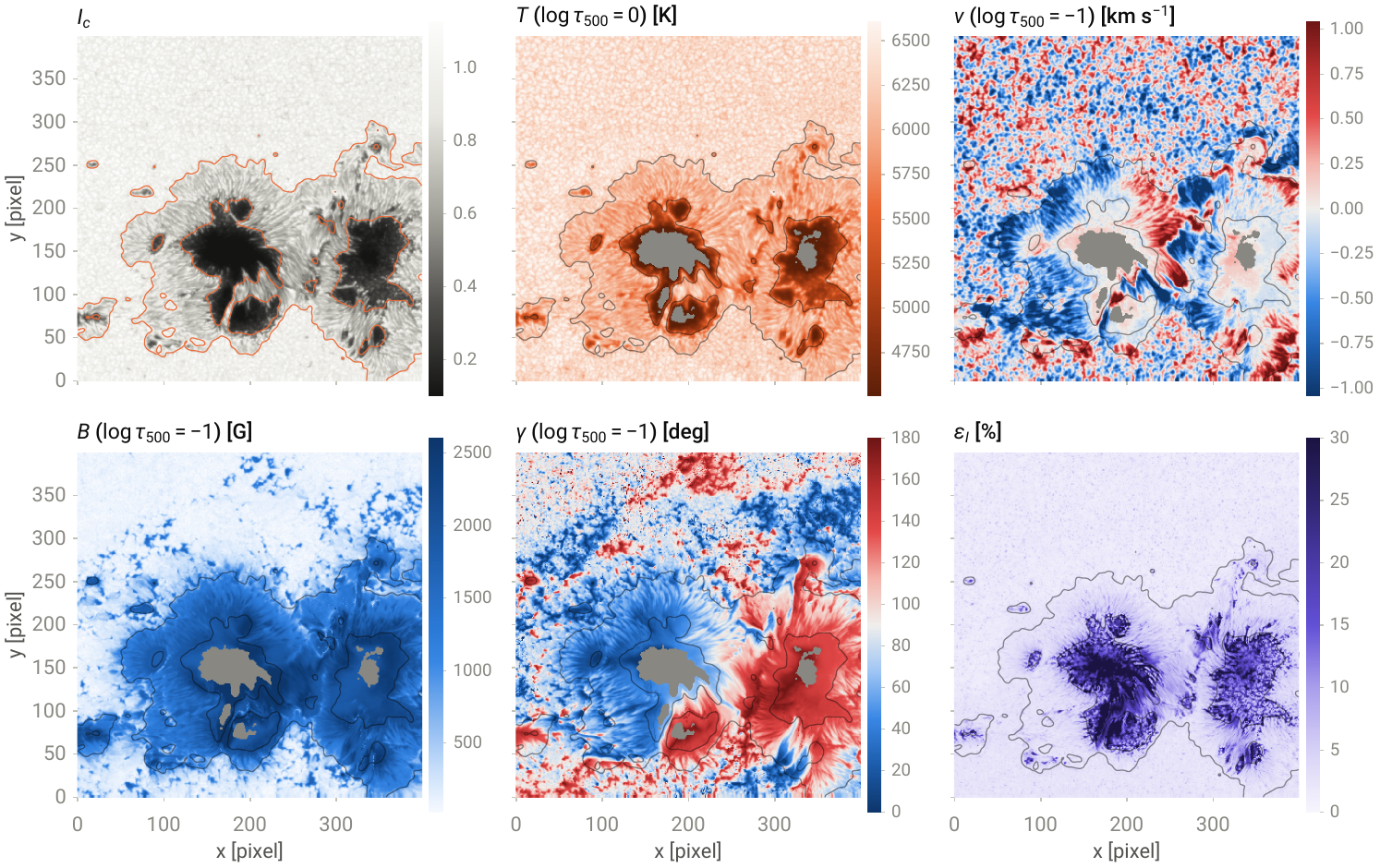}
\caption{Inversion of the Hinode SOT/SP observations of AR 11967. The upper six panels show the results of the 
inversion with PSF correction, while the lower six panels show the results of the inversion without PSF correction. 
From left to right and top to bottom: continuum intensity,
temperature at $\log \tau_{500}=0$, LOS velocity at $\log \tau_{500}=-1$ (positive values are redshifts),
field strength and inclination with respect to the LOS at $\log \tau_{500}=-1$, and relative rms residual of Stokes $I$, $\varepsilon_I$.
The gray areas in the maps of the physical quantities mask the coolest umbral regions, with a continuum intensity below 15\% of that of the
quiet Sun, where the inferred values are not reliable. The contours mark the
50\% and 90\% levels of the quiet-Sun continuum intensity (smoothed), which define the umbra, penumbra, and surroundings.
\label{fig:hinode_maps}}
\end{figure*}
        
\subsection{Initialization}
\label{sec:initialization}
Although the regularization helps the optimization avoid local minima, a reasonable initialization
accelerates convergence significantly. The temperature and the microturbulent velocity are initialized
with the median stratification of the training set. The LOS velocity is initialized with the velocity
derived from the center of gravity of Stokes $I$ of the photospheric lines. The magnetic field vector is initialized
with the least-squares estimation of the weak-field approximation \citep[e.g.,][]{2012MNRAS.419..153M}.
For the He\,{\sc i} slab, the velocity is initialized with the center of gravity of the red component
of the multiplet, the optical depth with the depth of the red component with respect to the nearby continuum, and the magnetic field with
the weak-field approximation applied to the red component, rotated to the local vertical reference frame.
The Doppler width and the damping are initialized to typical values of 8~km~s$^{-1}$ and 0.3, respectively, 
which are, in our experience, typical values in many observations \citep[see, e.g.,][]{2008ApJ...683..542A}.
When several slabs are used, the optical depth is divided among them and their velocities are slightly displaced, so that the
slabs do not start from identical solutions.

\section{Illustrative examples}
\label{sec:examples}
We illustrate the method with the inversion of three very different data sets: a Hinode SOT/SP map of an active region
in the Fe\,{\sc i} doublet, which constrains only the photosphere, a GRIS map in the 10830~\AA\ region, in
which the photosphere and the chromosphere are inferred simultaneously, and finally a CRisp Imaging Spectropolarimeter
\citep[CRISP;][]{2006A&A...447.1111S, 2008ApJ...689L..69S, 2026A&A...705A..55S} map in the 8542~\AA\ region. In all cases, all observed pixels are very close to 
disk center. 

\subsection{Hinode SOT/SP observations}
\label{sec:hinode}
The first data set is a $400 \times 400$ pixel subfield of a map of the active region NOAA 11967 obtained with the
spectropolarimeter \citep{2001ASPC..236...33L} of the Solar Optical Telescope \citep{2008SoPh..249..167T} onboard 
Hinode \citep{kosugi_hinode07}. The observations were taken on 4 February 2014 starting at 19:00~UT and ending at 19:52~UT, with 
a pixel size of 0.32\arcsec\ along the slit and 0.29\arcsec\ along the scan direction, and an exposure time of 4.8~s per 
slit position. The region is located at an heliocentric
angle of $\theta=14.5^\circ$ ($\mu=0.97$). This region is chosen as a benchmark for its enormous complexity, which
can put the emulators to the test.

\begin{figure*}
\centering
\includegraphics[width=0.85\textwidth]{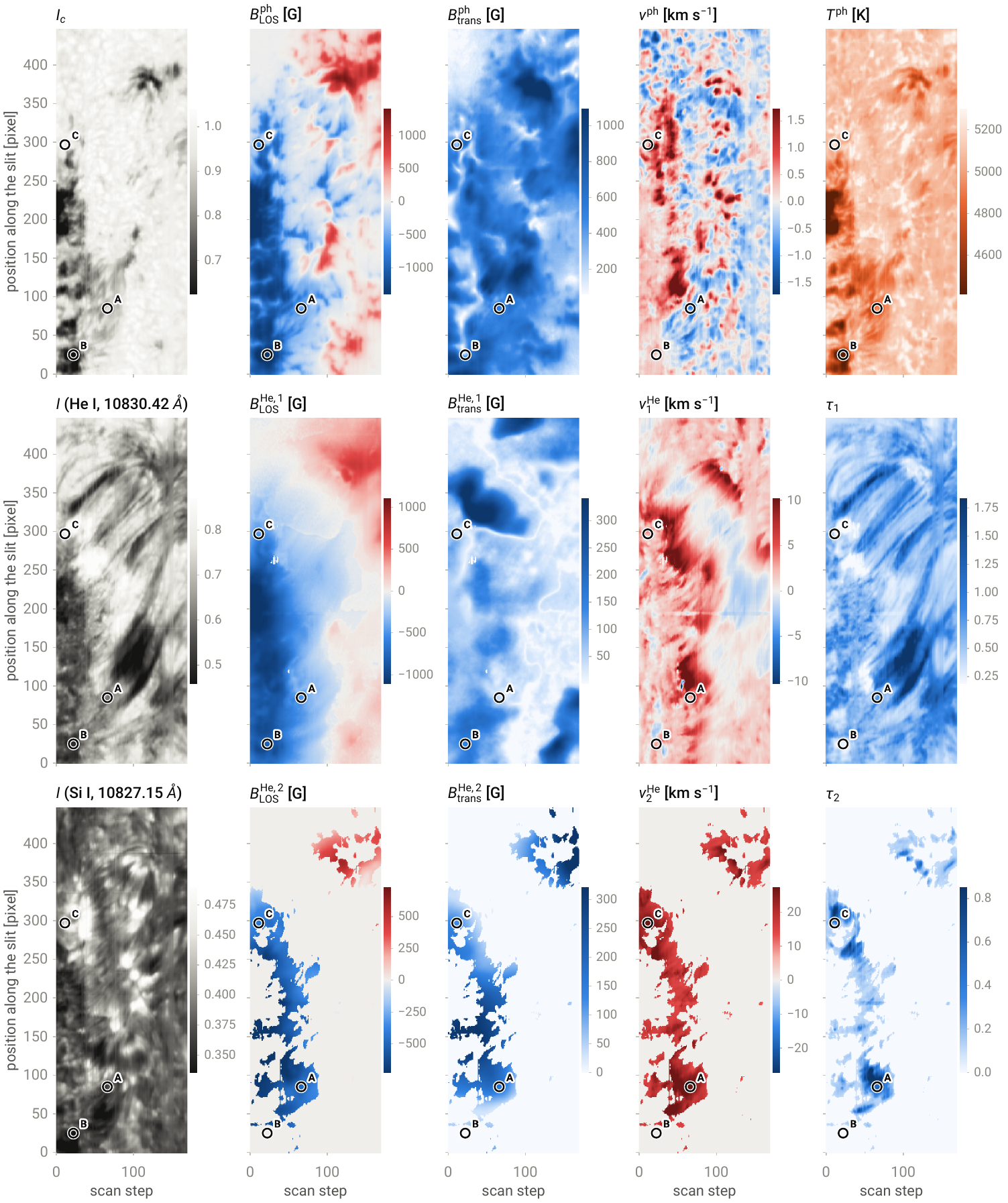}
\caption{Inversion of the GRIS observations in the 10830~\AA\ region. The first column, from top to bottom, shows monochromatic
images in the continuum, core of the red component of the He\,{\sc i} multiplet, and core of the Si\,{\sc i} line. 
The second, third and fourth columns show the inferred longitudinal and transverse components of the magnetic field, and the 
LOS velocity for the photospheric and chromospheric components. The rightmost column shows
the photospheric temperature at $\log \tau_{500}=-1$ and the optical depths of the chromospheric
components. Velocities are referred to the median photospheric velocity, with positive values indicating redshifts. The 
circles mark the pixels shown in Fig.~\ref{fig:gris_analysis}.}
\label{fig:gris_maps}
\end{figure*}
    
\begin{figure*}
\centering
\includegraphics[width=\textwidth]{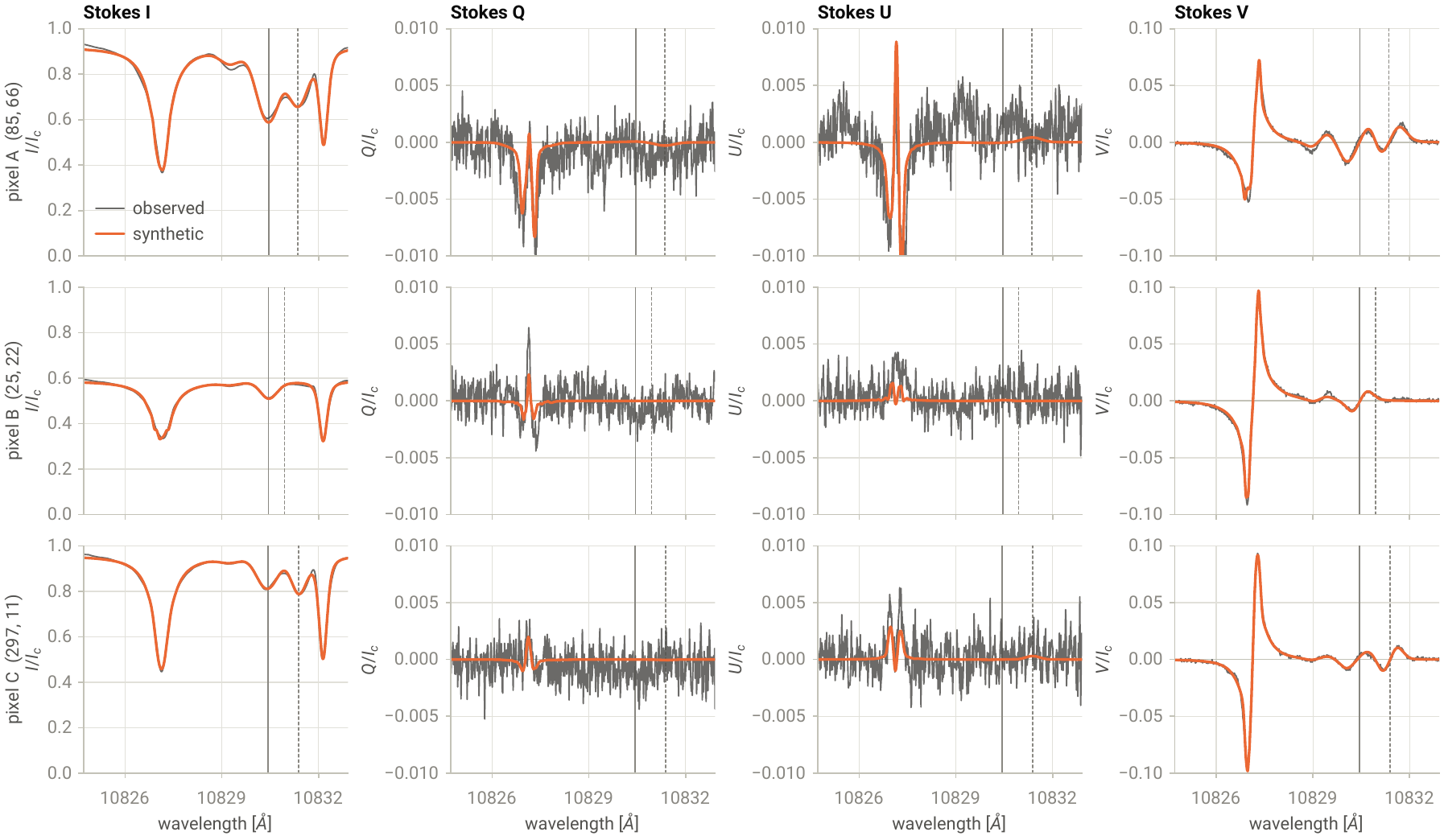}
\caption{Observed (black) and best-fit (orange) Stokes profiles of the three 
pixels marked in Fig.~\ref{fig:gris_maps}. The vertical lines indicate the inferred velocity shift of the
He\,{\sc i} red component of the first (solid) and second (dashed) slabs.}
\label{fig:gris_analysis}
\end{figure*}
    
The Stokes profiles are sampled in 112 wavelength points and are normalized to the
average continuum intensity of a quiet-Sun region. The FoV contains two large sunspots of opposite polarity, with 
several umbral cores,
together with pores and plage. We estimate the noise standard deviation per Stokes parameter to be 
$\sigma_I = 1.5 \times 10^{-3}$, $\sigma_Q = 1.6 \times 10^{-3}$,
$\sigma_U = 1.5 \times 10^{-3}$, and $\sigma_V = 2.9 \times 10^{-3}$. The value of $\sigma_I$ is smaller than 
the rms error of the Fe\,{\sc i} emulators in Stokes $I$. Following standard practice, we artificially augment 
it to $7.0 \times 10^{-3}$, which is equivalent to giving a larger weight to the Stokes $Q$, $U$, and $V$ components. 
The inversion is carried out with and without PSF correction during 1000 iterations.
The computing time per iteration on an NVIDIA 4090 GPU is 0.37~s without PSF correction and 0.42~s with 
PSF correction. This means that the computation time per pixel of the synthesis and computation of the
gradient is 2.3~$\mu$s without PSF correction and 2.6~$\mu$s with PSF correction. 

The photospheric model uses six nodes in temperature 
(at $\log \tau_{500} = -7.5$, $-4$, $-2.5$, $-1$, 0.0, and $1.5$), two
nodes for the LOS velocity and for each Cartesian component of the magnetic field (at $\log \tau_{500} = -7.5$ and $1.5$), resulting
in linear functions of $\log \tau_{500}$, and a height-independent microturbulent velocity.
We found that it is important to add a small amount of Tikhonov regularization if the effect of the
PSF is included in the forward model. Otherwise, a checkerboard pattern often appears in the maps of the physical parameters. 
This is a typical consequence of the fact that the PSF couples neighboring pixels, so that the inversion can trade off the parameters of adjacent pixels to produce the same degraded profiles.

Figure~\ref{fig:hinode_maps} shows the maps of some of the inferred physical quantities for the two
cases. The six upper panels correspond to the inversion with PSF correction, while the six lower panels 
correspond to the inversion without PSF correction. We display contours separating the umbra, penumbra, and surroundings 
using simple thresholds at 50\% and 90\% of the quiet-Sun continuum intensity. 
We show the continuum intensity, the temperature at 
$\log \tau_{500}=0$, the LOS velocity at $\log \tau_{500}=-1$ (referred to the median velocity of the umbra, 
with positive values being redshifts), the field strength and inclination with respect to the 
LOS at $\log \tau_{500}=-1$, and the relative rms residual of Stokes $I$,
$\varepsilon_I = \langle (I^\mathrm{syn}-I^\mathrm{obs})^2 \rangle_\lambda^{1/2} / \langle I^\mathrm{obs} \rangle_\lambda$.
We prefer $\varepsilon_I$ to the $\chi^2$ as a measure of the quality of the fit in these maps because the noise levels are fixed in units of
the quiet-Sun continuum, so that even clearly bad fits have a small $\chi^2$ in the darkest regions. Artifacts in the inferred maps
are seen in the regions of strongest magnetic fields, a consequence of the fact that these magnetic fields
are outside the ranges used in the emulators, something that can be easily solved by retraining them. Additional 
artifacts are present in the coolest regions of the umbra, where many molecular
lines appear, which cannot be reproduced by our synthesis model. We mask the regions with a continuum intensity
below 15\% of that of the quiet Sun, although $\varepsilon_I$ shows that the fit of Stokes $I$ is poor in most of the umbra.
Arguably, the most conspicuous issue
is the region of large residuals above the light bridge. This region is known to harbor one of the strongest 
magnetic fields ever measured in the photosphere \citep{2020ApJ...895..129C}.


\subsection{GRIS observations in the 10830~\AA\ region}
\label{sec:gris}
The second data set is a map of $447 \times 170$ pixels (along the slit and scanning direction, respectively) observed 
on 15 July 2024 with the GRIS
spectropolarimeter at GREGOR. The observed active region is NOAA 13753
with a pixel size of 0.135\arcsec\ and a spectral sampling of 18.3~m\AA\, px$^{-1}$. Dark and flat-field corrections were performed following
the standard procedures described by \cite{1999ASPC..184....3C,2003SPIE.4843...55C}.
The spectral window contains 559 wavelength
points from 10824.7~\AA\ to 10832.9~\AA. The FoV contains the edge of a sunspot of negative polarity on the left, a positive 
polarity pore at the top, and plage
in between (see the photospheric panels of Fig.~\ref{fig:gris_maps}). Several pixels display He\,{\sc i} profiles 
with two clearly separated components, so the
chromosphere is modeled with two He\,{\sc i} slabs placed in series. The height of both slabs is fixed 
to $h=3\arcsec$. In an effort to separate both slabs, we limit the range of velocities 
of the first component to $[-40, 40]$~km~s$^{-1}$ and of the second component to $[15, 40]$~km~s$^{-1}$. The
photosphere is modeled with six nodes in temperature (at $\log \tau_{500} = -7.5$, $-4$, $-2.5$, $-1$, 0.0, and $1.5$), two nodes 
for the LOS velocity and $B_z$, and one node for $B_x$ and $B_y$. Finally, following the
standard practice in \texttt{Hazel2}, the telluric line is modeled with a Voigt
profile whose position, width, depth, and damping are free in each pixel. All photospheric parameters are slightly Tikhonov regularized
to avoid pixel-to-pixel inversion noise. The same applies to the magnetic field parameters of the chromosphere. Finally,
the optical depth of the second He\,{\sc i} slab is regularized 
with an $\ell_1$ norm to favor a small number of pixels with large opacities. We tune the weight of the regularization until 
the desired level of sparsity produces a second slab to be present only in the pixels where it is really needed. A value of
$\lambda=1000$ does a good job but could also be tuned in a more systematic way. 

\begin{figure*}
\centering
\includegraphics[width=\textwidth]{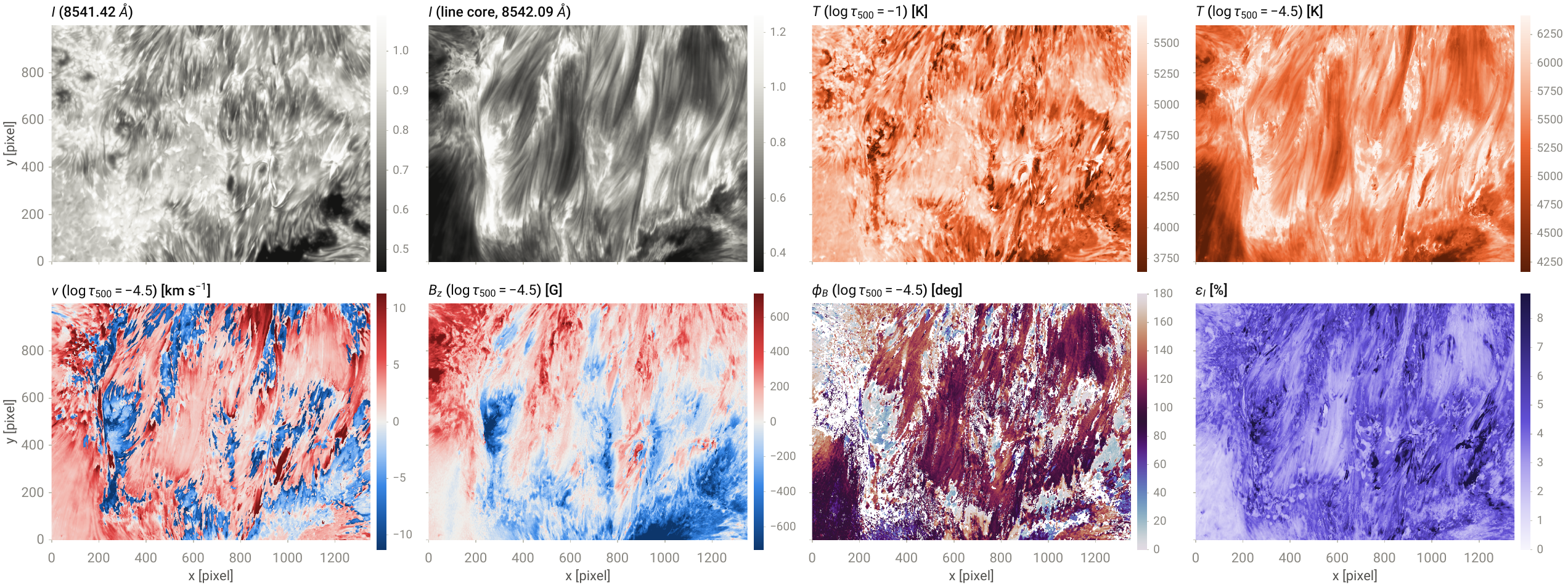}
\caption{Inversion of the CRISP observations. The upper four panels show the wing and line core
intensities, the temperature at $\log \tau_{500}=-1$ and at $\log \tau_{500}=-4.5$.
The lower four panels display the LOS velocity at $\log \tau_{500}=-4.5$ (positive values are redshifts),
the longitudinal and the azimuth of the field at $\log \tau_{500}=-4.5$, and the relative rms residual of Stokes $I$, $\varepsilon_I$.
The Stokes $Q>0$ reference direction, and origin of azimuths, is the horizontal direction.}
\label{fig:crisp_maps}
\end{figure*}

The inversion is run for 500 iterations, although practical convergence is reached after 200 iterations.
The computing time per iteration on an NVIDIA 4090 GPU is 2.46~s, resulting in 32.4~$\mu$s per pixel for
the synthesis and computation of the gradient. Simultaneous inversion of the telluric line turns out to be very useful. 
The position of the telluric line drifts 
almost linearly during the scan by 61~m\AA, equivalent to 1.7~km~s$^{-1}$, which can be associated with a systematic
shift of the spectrograph. We subtract the drift from all inferred velocities so that all velocities 
are referred to the median photospheric velocity of the map.

Figure~\ref{fig:gris_maps} shows the main results,
with Fig.~\ref{fig:gris_analysis} showing the observed and synthetic 
Stokes profiles of the three marked pixels in Fig.~\ref{fig:gris_maps}. As context, Fig.~\ref{fig:gris_maps} shows 
monochromatic images in the continuum, core of the red component of the He\,{\sc i} multiplet, and core of 
the Si\,{\sc i} line. Pixel A and C are representative of
regions with supersonic downflows, while pixel B is representative of the fit on the sunspot. The fits are very good 
in all cases.

The optical depth and velocity maps of the first He\,{\sc i} slab
displays the fibrillar structure that is typical of active region chromospheres. 
Despite the large spatial structure in these quantities, the inferred magnetic field is 
rather homogeneous. The line-of-sight component of the magnetic field is well constrained by the
presence of sizable Stokes $V$ signals in the He\,{\sc i} lines. On the contrary,
the Stokes $Q$ and $U$ signals in the He\,{\sc i} lines are very weak, so the azimuth of the 
magnetic field is dominated by the regularization and the initial guess. We caution that the
transverse component of the magnetic field is not well constrained in this data set, and that the 
inferred values should be taken with caution.

The second slab
has an optical depth larger than 0.1 in only 17\% of the pixels. Pixels in which the optical depth does 
not reach these values are masked in the lower panels of Fig.~\ref{fig:gris_maps}. 
The first slab, with a median optical depth of 0.59, is almost at rest, with a median 
velocity of 1.5~km~s$^{-1}$. The second slab is systematically redshifted, with all pixels with velocities 
exceeding 10~km~s$^{-1}$, and 11\% of the pixels with velocities larger than 20~km~s$^{-1}$. 
These pixels define the patches of strong downflows visible in 
Fig.~\ref{fig:gris_maps}, which are also characterized by similar Doppler widths 
(9~km~s$^{-1}$). Supersonic downflows in the He\,{\sc i} 10830~\AA\ triplet have 
been reported in a variety of chromospheric structures, including arch filament 
systems (AFSs) and emerging flux regions \citep{Solanki2003b, Aznar2007, Lagg2007, Xu2010, 2019A&A...625A.128D, GonzalezManrique2016, GonzalezManrique2018,  GonzalezManrique2020}. The similar Doppler
widths suggest that these supersonic downflows are fully resolved and are not producing a 
significative increase in the plasma temperature. 


\subsection{CRISP observations in the 8542~\AA\ region}
\label{sec:crisp}
The final example is a 1300 $\times$ 900 pixels region that is part of a larger area reconstructed from an observation of AR13354 recorded on 29 June 
2023 at 09:05UT at solar coordinates $(X,Y)$=$(100",200")$. The data were acquired in the \ion{Ca}{ii}~8542~\AA\ line, 
sampled at 15 line positions with the CRISP Fabry-P\'erot spectropolarimeter at the Swedish 1-m Solar Telescope. The 
data were reduced and calibrated using the SSTRED data pipeline \citep{2021A&A...653A..68L,2015A&A...573A..40D}. 
Atmospheric seeing effects were minimized using the multi-object multi-frame blind deconvolution code 
\citep[MOMFBD,][]{vannoort05}. We estimate $\sigma_I = 1.0 \times 10^{-2}$, $\sigma_Q = 2.5 \times 10^{-3}$,
$\sigma_U = 2.0 \times 10^{-3}$, and $\sigma_V = 2.0 \times 10^{-3}$, although the
estimation in Stokes $I$ is difficult given that the observations do not reach
the continuum wavelengths. This dataset was recently used by \citet{2025A&A...696A...3L} and we refer 
the reader to that publication for more details.

The atmosphere is modeled with seven nodes in temperature (at $\log \tau_{500} = -7.5$, $-5.5$, $-4$, $-2.5$, $-1$, 0.0, and $1.5$), three nodes 
for the LOS velocity, and two nodes for $B_x$, $B_y$, $B_z$ and the microturbulent velocity. We regularize all 
components of the magnetic field, given that the polarimetric signal (especially the linear polarization) is 
relatively noisy.

The inversion is run for 500 iterations with a total computing time of 15 min. 
The computing time per iteration on an NVIDIA 4090 GPU is $\sim$2~s, resulting in 1.4~$\mu$s per pixel for
the synthesis and computation of the gradient. This calculation is split in five batches to
saturate the available VRAM.

The results of the inversion are shown in Fig. \ref{fig:crisp_maps}. The FOV includes a 
complex active region with a chromosphere dominated by elongated fibrils along the $y$-axis. The 
magnetic field at the footpoints of these fibrils is strongly vertical. We find very strong ($\sim$10 km\, s$^{-1}$)
and localized redshifts (e.g., coordinates $(x,y)=(200,600)$) that are aligned with the dark fibrils while strong blueshifts (also of the same
order of magnitude) are found just a few arcsec apart, everything lying along a polarity inversion line.
Although the Stokes $Q$ and $U$ profiles are noisy in a significant fraction of the pixels, the azimuth of the magnetic
field displays a clear alignment with the fibrils \citep[see, e.g.,][]{2017A&A...599A.133A}.

\section{Conclusions}
\label{sec:conclusions}
We have shown that neural emulators make spatially
regularized inversions of full fields of view fast enough to be routinely applied. The neural 
network surrogate, which is differentiable by construction, can be run in parallel for many 
pixels on a GPU. Unlike direct neural inverters, the observations are
still fitted through a physical forward model, so that the user retains the freedom to choose the
number and location of the nodes, the number of atmospheric components, and the regularization of each
physical parameter, all without retraining.

The inversion is formulated as a single optimization problem for all pixels, which is solved with a
proximal gradient method that handles non-differentiable regularizations such as total variation
or sparsity in the wavelet and pixel domains. The separability of the data term allows us to process the field of
view in chunks, so that arbitrarily large maps can be inverted on a single GPU. When the spatial PSF is included,
a two-pass scheme provides the exact gradient at the cost of an additional forward pass. Large maps
like the ones presented here can be inverted in minutes, reducing the turn-around time for investigating
node placement, regularization, and the number of components.

The examples shown here also expose the main limitations of the approach. Emulators can only be trusted inside the region
of the parameter space covered by the training set. Regions with very strong magnetic fields, such as the
light bridge of AR 11967, or with very strong velocities, like footpoints of the AFSs, lie at the edge 
or outside this region and display artifacts and large values of $\varepsilon_I$.
Likewise, physical ingredients that are absent from the training set, such as the molecular lines
of the coolest umbral regions, cannot be reproduced. All these issues can be easily mitigated by extending the training set
to more extreme atmospheres and by including the missing physics in the synthesis, at the price of
retraining the emulators. Additionally, the emulators only provide an approximation to a certain
level of precision. We find that this level is currently enough for extracting relevant physical information
from the observations. If more precision is needed, iterative refinement can be considered, where
successive neural emulators are trained to correct the residuals between the synthetic profile and the 
emulated one \citep{WANG2024112865}.

The approach is general and can be extended in several directions. Since the cost of the synthesis is
paid only once during the generation of the training set, emulators for lines formed under
non-LTE conditions with partial frequency redistribution effects (either with polarization or not), such as the 
Ca\,{\sc ii} H and K lines or the Mg\,{\sc ii} h and k lines, would
permit spatially coupled inversions of chromospheric lines that are currently very time consuming to perform.
Combining emulators of several spectral regions observed simultaneously is straightforward, as is the
inclusion of spatially variant PSFs or observations with different spatial resolutions.
Additionally, these emulators can also be coupled with neural fields \citep{2025A&A...693A.170D,2025ApJ...985L...7J}
to enforce consistency with physical constrains. We believe that this approaches will become an important 
tool for the routine analysis of spectropolarimetric observations.

\section*{Data availability}

The code is publicly available at \urlcode.

\begin{acknowledgements}
The authors acknowledge assistance from Claude Opus 5.5 during the development of the emulators and
the inversion code, as well as polishing the manuscript. We also acknowledge 
partial funding from the Agencia Estatal de Investigación del Ministerio de Ciencia, Innovación 
y Universidades (MCIU/AEI) under grant 
``Polarimetric Inference of Magnetic Fields'' and the European Regional Development Fund (ERDF) with 
reference PID2022-136563NB-I00/10.13039/501100011033, as well as grants PID2024-156066OB-C55 and PID2024-156538NB-I00, both funded by 
MCIN/AEI/10.13039/501100011033 and by ``ERDF A way of making Europe''. S.J.G.M. acknowledges grant RYC2022-037565-I and C.K. 
grant RYC2022-037660-I, both funded
by MCIN/AEI/10.13039/501100011033 and by “ESF Investing in your future”.
This work is also part of grant CEX2025-001609-S (Severo Ochoa Centre of Excellence program), awarded to the Instituto 
de Astrof\'{\i}sica de Canarias and funded by MICIU/AEI/10.13039/501100011033.
This research was partially supported by the Research Council of Norway through its Centres of Excellence scheme, project number 262622.
This project has been partially funded by the European Union through the European Research Council (ERC) under the Horizon Europe program (MAGHEAT, grant agreement 101088184).
The computations was enabled by resources provided by the National Academic Infrastructure for Supercomputing in Sweden (NAISS), partially funded by the Swedish Research Council through grant agreement no. 2022-06725, in particular at the PDC Centre for High Performance Computing (PDC-HPC) at the Royal Institute of Technology in Stockholm within the projects NAISS 2025/1-9 and 2026/1-20. The Institute for Solar Physics is supported by a grant for research infrastructures of national importance from the Swedish Research Council (registration number 2021-00169). The Swedish 1-m Solar Telescope is operated on the island of La Palma by the Institute for Solar Physics of Stockholm University in the Spanish Observatorio del Roque de los Muchachos of the Instituto de Astrof\'isica de Canarias.
This research has made use of NASA's Astrophysics Data System Bibliographic Services.
We acknowledge the community effort devoted to the development of the following open-source packages that were
used in this work: \texttt{numpy} \citep[\texttt{numpy.org},][]{numpy20}, 
\texttt{matplotlib} \citep[\texttt{matplotlib.org},][]{matplotlib}, \texttt{PyTorch} 
\citep[\texttt{pytorch.org},][]{pytorch19}, \texttt{scipy} \citep[\texttt{scipy.org},][]{2020SciPy-NMeth} and \texttt{scikit-learn} \citep[\texttt{scikit-learn.org},][]{scikit-learn}.
\end{acknowledgements}

%
\bibliographystyle{aa} 
\bibliography{biblio} 
%

\appendix
\section{Regularizers}
\label{sec:appendix_regularization}
The following regularizers are the ones currently implemented in the code:
\begin{itemize}
\item Tikhonov (also known as $\ell_2$) regularization of the spatial gradient, $R(\vect{X})=\|\nabla \vect{X}\|_2^2$,
where $\nabla$ is computed with forward differences. It favors
smooth maps and penalizes quadratically the jumps between neighboring pixels.
We note that a regularization of the $B_x$ and $B_y$ components of the magnetic field is not enough 
to guarantee a smooth field when $B_x$ is close to zero. In order to avoid artifacts, we therefore 
regularize the non-ambiguous components $B_t \cos 2 \phi$ and $B_t \sin 2 \phi$, where $B_t=\sqrt{B_x^2+B_y^2}$ is 
the transverse field strength, while enforcing $B_x \geq 0$.
\item Isotropic total variation \citep[TV;][]{rudin1992nonlinear}, $R(\vect{X})=\sum_{ij} \sqrt{(\nabla_x X)_{ij}^2+(\nabla_y X)_{ij}^2}$.
It favors piecewise constant maps and, contrary to the Tikhonov regularization, it preserves sharp
discontinuities.
\item The $\ell_1$ norm of the map, $R(\vect{X})=\|\vect{X}\|_1$, which favors maps in which most of the pixels
are zero. It can be useful, for instance, for the magnetic field in quiet Sun regions, or for multi-component inversions.
\item The $\ell_1$ norm of the detail coefficients of an orthonormal 2D wavelet transform $\mathbf{W}$,
$R(\vect{X})=\|\mathbf{D}\mathbf{W}\vect{X}\|_1$, where $\mathbf{D}$ selects the
detail coefficients (the approximation coefficients at the coarsest scale are not penalized).
This is the sparsity regularization used by \cite{2015A&A...577A.140A}, and it favors maps that can be described with a few
wavelet coefficients, a property that solar images fulfill to a large extent
\citep[e.g.,][]{starck06}. We use periodic transforms built from the orthogonal
wavelets of \cite{daubechies1992ten}, with a few decomposition levels.
\end{itemize}

Given that the regularization terms might be not differentiable, they are handled using proximal gradient methods
\citep{ista09,parikh2013proximal}. Each regularization term is enforced by applying the appropriate proximal projection operator, defined as
\begin{equation}
\mathrm{prox}_{\gamma R}(\vect{X}) = \argmin_{\vect{Z}} \left[ \frac{1}{2} \| \vect{Z} - \vect{X} \|_2^2 + \gamma R(\vect{Z}) \right],
\end{equation}
where $\gamma$ is a step size. The proximal operator acts as a denoiser that pulls the maps towards the
solutions favored by the regularization. All regularization functions considered in this work have proximal operators that can
be computed exactly or very efficiently. For the $\ell_1$ norm, it is the soft-thresholding operator,
$\mathrm{prox}_{\gamma \|\cdot\|_1}(\vect{X}) = \mathrm{sign}(\vect{X}) \max(|\vect{X}|-\gamma, 0)$, applied
element-wise. Since $\mathbf{W}$ is orthonormal, the proximal operator of the wavelet regularization is
obtained by soft-thresholding the detail coefficients, $\mathbf{W}^T S_\gamma(\mathbf{W}\vect{X})$.
The proximal operator of the Tikhonov term requires the solution of the linear system
$(\mathbf{1}+2\gamma \nabla^T \nabla)\vect{Z}=\vect{X}$. With appropriate boundary conditions, the discrete
Laplacian $\nabla^T \nabla$ can be diagonalized by the 2D discrete cosine transform, so the solution is computed
exactly with two DCTs and a division by $1+2\gamma(e_k+e_l)$, where $e_k=2-2\cos(\pi k/N)$ are the
eigenvalues of the 1D Laplacian. Finally, the proximal operator of the TV term has no closed form, and is computed
with the fast gradient projection algorithm of \cite{beck2009fast} applied to the dual problem, with a
fixed number of 50 iterations.
If a certain node map has several
regularization terms, their proximal operators are applied sequentially. Finally, the parameters are 
projected onto their bounds using the proximal operator of the box constraints. 

\section{Gradients when doing PSF deconvolution}
\label{sec:appendix_psf}
As already described in the main text, when the spatial PSF of the telescope is included in the forward model,
we need to replace the Stokes parameters with
$\tilde{S}^\mathrm{syn}_{s,p}(\lambda_i)=[\mathbf{P} * S_s^\mathrm{syn}(\lambda_i)]_p$, making the problem
non-separable in independent pixels. However, we note that the coupling enters 
only through the linear operator $\mathbf{P}$ while the synthesis remains
fundamentally local. Applying the chain rule gives:
\begin{equation}
\frac{\partial \chi^2}{\partial \vect{x}_p} = \frac{\partial \chi^2}{\partial {\vect{S}}^\mathrm{syn}_p} 
\frac{\partial \vect{S}^\mathrm{syn}_p}{\partial \vect{x}_p}
\label{eq:psf_gradient1}
\end{equation}
where
\begin{equation}
\frac{\partial \chi^2}{\partial {\vect{S}}^\mathrm{syn}_p} =
\frac{\partial \chi^2}{\partial {\tilde{\vect{S}}}^\mathrm{syn}_p} 
\frac{\partial \tilde{\vect{S}}^\mathrm{syn}_p}{\partial {\vect{S}}^\mathrm{syn}_p}=
\mathbf{P}^T * \frac{\partial \chi^2}{\partial \tilde{\vect{S}}^\mathrm{syn}_p},
\label{eq:psf_gradient2}
\end{equation}
where we have used the fact that the derivative of a convolution is the convolution with the flipped kernel. 
Equation (\ref{eq:psf_gradient2}) can be computed without storing the intermediate activations of the emulators, 
since it only involves a convolution with the PSF and the quadratic term of the $\chi^2$.
This gradient is multiplied by the Jacobian of the emulators to compute the final gradient.

From a practical perspective, the gradient of Eq.~(\ref{eq:psf_gradient1}) can be computed with a two-pass scheme. 
In the first pass, the undegraded profiles are synthesized chunk by chunk without
building the computational graph. The synthetic profiles are then used as a leaf tensor\footnote{A leaf tensor is a tensor that is 
created at the very start of the computational graph.},
convolved with the PSF, storing the computational graph with 
automatic differentiation (in this case, only the cheap convolution). In the second pass, every chunk is 
synthesized again, this time keeping the computational graph, and using the previously computed gradients as 
input, finally computing the
vector-Jacobian product of Eq.~(\ref{eq:psf_gradient1}). The price to pay is an additional forward pass but, given
that the computation of the gradient is the most expensive part of the process, this overhead is acceptable. When the FoV fits in a
single chunk, the graph of the first pass is kept and a single forward and backward pass is carried out.

\end{document}